\documentclass[11pt]{article}
\usepackage{hyperref}
\pdfoutput=1

\begin{document}
\title{Spontaneous Jumping of Coalescing Drops on a Superhydrophobic Surface}
\author{Jonathan B. Boreyko and Chuan-Hua Chen \\
\\\vspace{6pt} Department of Mechanical Engineering and Materials Science, \\ Duke University, Durham, NC 27708, USA}
\maketitle

Fluid dynamics video

\begin{abstract}
When micrometric drops coalesce in-plane on a superhydrophobic surface, a surprising out-of-plane jumping motion was observed. Such jumping motion triggered by drop coalescence was reproduced on a Leidenfrost surface. High-speed imaging revealed that this jumping motion results from the elastic interaction of the bridged drops with the superhydrophobic/Leidenfrost surface. Experiments on both the superhydrophobic and Leidenfrost surfaces compare favorably to a simple scaling model relating the kinetic energy of the merged drop to the surface energy released upon coalescence. The spontaneous jumping motion on water repellent surfaces enables the autonomous removal of water condensate independently of gravity; this process is highly desirable for sustained dropwise condensation.
\end{abstract}

\href{http://ecommons.library.cornell.edu/bitstream/1813/14095/2/APS_DFD_small.mp4}{Video
1}.

\end{document}